\documentclass[final,5p,times,twocolumn]{elsarticle}

\usepackage{hyperref}
\usepackage{lineno}
\usepackage{graphicx}
\usepackage{amsmath}
\usepackage{color}
\usepackage{verbatim}
\usepackage[utf8]{inputenc}
\usepackage{memhfixc}
\usepackage{bold-extra}

\journal{Nuclear Instruments and Methods B}

\begin{document}


\begin{frontmatter}

\title{Bayesian data analysis tools for atomic physics}

\author{Martino Trassinelli}
\address{Institut des NanoSciences de Paris, CNRS, Sorbonne Universit\'es, UPMC Univ Paris 06, F-75005, Paris, France}
\ead{trassinelli@insp.jussieu.fr}

\begin{abstract}
We present an introduction to some concepts of Bayesian data analysis in the context of atomic physics.
Starting from basic rules of probability, we present the Bayes' theorem and its  applications.
In particular we discuss about how to calculate simple and joint probability distributions and the Bayesian evidence, a model dependent quantity that allows to assign probabilities to different hypotheses from the analysis of a same data set.
To give some practical examples, these methods are applied to two concrete cases.
In the first example, the presence or not of a satellite line in an atomic spectrum is investigated.
In the second example, we determine the most probable model among a set of possible profiles from the analysis of a statistically poor spectrum.
We show also how to calculate the probability distribution of the main spectral component without having to determine uniquely the spectrum modeling.
For these two studies, we implement the program \texttt{Nested\_fit} to calculate the different probability distributions and other related quantities.
\texttt{Nested\_fit} is a Fortran90/Python code developed during the last years for analysis of atomic spectra. 
As indicated by the name, it is based on the nested algorithm, which is presented in details together with the program itself.
\end{abstract}


\end{frontmatter}

\section{Introduction}
Commonly, a data analysis is based on the comparison between a function $F(\boldsymbol{a})$ used to model the data that depends on a set of parameters $\boldsymbol{a}$ (ex. $a_1 \to $ amplitude, $a_2 \to $ position, etc.) and the data them-self that consist in recorded number of counts $y_i$ at each channel $x_i$.
The estimation of the parameter values describing at best the data is generally obtained by the maximum likelihood method (and its lemma, the method of the least squares), which consists to find the values  $\boldsymbol{a}^{best}$ that maximize the product of the probabilities for each channel  $x_i$  to observe $y_i$ counts for a given expected value $F(x_i,\boldsymbol{a}^{best})$.

Even if very successfully in many cases, this method has some limitations. 
If some function parameter is subject to constraints on its values (as ex. one model parameter could be a mass of a particle, which cannot be negative), the corresponding boundary conditions cannot be taken into account in a well defined manner. 
With the likelihood function we are in fact calculating the probabilities to observe the data $\{x_i,y_i\}$  for given parameter values and not the probability to have certain parameter values for given experimental data.

An additional difficulty for the maximum likelihood method arises when different hypotheses are compared, represented for example by two possible modeling functions $F_A$ and $F_B$, in view of the acquired data.
The determination of the most adapted model describing the data generally done with goodness-of-fit tests like the $\chi^2$-test, the likelihood, etc. \cite{Neyman1933,Yates1934,Bevington,Massey1951,Akaike1974,Spiegelhalter2002}.
In the unfortunate case where there is no clear propensity to an unique model and we are interested on the value of a parameter common to all models (as the position of the a peak with undefined shape), no sort of weighted average can be computed from goodness-of-fit test outcomes. 
To do this, the assignment of a probability $P(\mathcal{M})$ to each model is mandatory, which cannot be calculated in the classical statistics framework.

Another important and fundamental problem of the common data analysis approach is the requirement of repeatability for the definition of \textit{probability} itself. 
In classic data analysis manuals we can find sentences as:
\begin{quotation}
\textit{``Suppose we toss a coin in the air and let it land. There is 50\% probability that it will land heads up and a 50\% probability that it will land tails up. By this we mean that if we continue tossing a coin repeatedly, the fraction of times that it lands with heads up will asymptotically approach 1/2 \dots}'' \cite{Bevington}
\end{quotation}
This definition is completely inadequate to rare processes as those encountered for example in cosmology, where several models are considered to describe one unique observation, our universe, and more recently in gravitational-wave astronomy, where at present only two observations are available \cite{Abbott2016a,Abbott2016b}.

To overcome these problems, a different approach has to be implemented with a new and more general definition of probability.
This approach is the result of the work of Th.~Bayes, P.-S.~Laplace, H.~Jeffreys and of many others \cite{Bayes1763,Laplace1825,Jeffreys,Jaynes} and is commonly called \textit{Bayesian statistics}.

Bayesian methods are routinely used in many fields: cosmology \cite{Lewis2002,Trotta2008,Feroz2009}, particle physics \cite{PDG2016}, nuclear physics, \dots.
In atomic physics their implementation is still limited (e.g. in atomic interferometry \cite{Stockton2007,Calonico2009}, quantum information \cite{Wiebe2016}, ion trapping \cite{Mooser2013}, ion--matter interaction \cite{Barradas1999}, etc.) with almost no use in atomic spectroscopy, even if in some cases it would be strongly required.
For example, when we want to determine the correct shape of a instrumental response function we are actually testing hypotheses, as in the case of the determination of the presence or not of possible line contributions in a complex or statistically poor spectrum.

The goal of this article is to present a basic introduction of Bayesian data analysis methods in the context of atomic physics spectroscopy and to introduce the program \texttt{Nested\_fit} for the calculation of distribution probabilities and related quantities 
(mean values, standard deviation, confidence intervals, etc.)
from the application of these methods.
The introduction to Bayesian statics is based in the extended literature on this domain, and in particular on Refs.~\citenum{Jeffreys,Trotta2008,Cox1946,Ballentine1986,Sivia}.
For a clear and practical presentation, we will present two simple applications of data analysis where we implemented a Bayesian approach using the \texttt{Nested\_fit} program. 
The first example is about the probability evaluation of the presence of a satellite peak in a simple atomic spectrum.
The second one deals with the analysis of a statistically poor spectrum in which one or multiple peaks contributions has to be considered and where possible aberrations in the response function have also to be taken into account.
We will in particular show how to assign probabilities to the different models from the experimental data analysis
and compare them to classical goodness-of-fit tests.
Moreover, we will see how to extract the probability distribution of the main peak position without the need to uniquely choose between the different models.

The article is organized as following.
A general definition of probability and Bayesian statistic concepts as the Bayes' theoreme and \textit{Bayesian evidence} are present in Sec.~\ref{sec:probability}, together with a very general and axiomatic definition of probability deduced from simple logic arguments.
In Section~\ref{sec:nested} we present in details the nested algorithm for the calculation of the Bayesian evidence and in Sec.~\ref{sec:nested_fit} we will see its implementation in the program \texttt{Nested\_fit}, which is also presented. 
These two sections are quite technical and they could be skipped in a first reading.
Section~\ref{sec:applications} is dedicated to the Bayesian data analysis applications to two real data sets and Sec.~\ref{sec:conclusion} is our conclusion.
Two appendixes are also proposed: one about the introduction of information and complexity concepts in the context of Bayesian statistics, and a second about the evaluation of the uncertainty of the Bayesian evidence calculated by the nested sampling method.

\section{Probability} \label{sec:probability}

\subsection{Probability axioms} \label{sec:axioms}
A very general definition of probability $P(X)$ can be obtained by trying to assign real numbers to a certain degree of plausibility or believe than assertions $X,Y$, etc., would be true.
$X$ and $Y$ assertions are very general. They can be assertions of specific statements (ex. ``In the next toss the coin will land heads'') or implying values (ex. the parameter $b$ is in a certain range $[b_{min},b_{max}$]).
When basic logic and consistency are required, the form of the probability $P$ is ensured by the axioms \cite{Cox1946,Cox,Jaynes,Sivia,Ballentine1986}
\begin{align}
&0\le P(X | I)\le 1, \label{eq:cox1} \\
&P(X | X,I)= 1, \label{eq:cox2} \\
&P(X | I) + P(\bar X | I) = 1, \label{eq:cox3} \\
&P(X,Y | I) = P(X | Y,I) \times P(Y | I). \label{eq:cox4}
\end{align}
In the equations above, $\bar X$ indicates the negation of the assertion $X$ (not-$X$);
the vertical bar ``$|$'' means ``given'' and where $I$ represents the current state of knowledge. 
For example, $I$ can represent the ensemble of the physics laws describing a certain phenomenon, e.g. the thermodynamics laws, and $X$, $Y$ can represent two quantitative measurements related to this phenomenon, e.g. two temperature measurements at different times of a cooling body.
The joint probability $P(X,Y | I)$ means that both ``$X$ AND $Y$'' are true (equivalent to the logical conjunction `$\wedge$'). 
The deduction of these axioms have been obtained for the first time in 1946 by Richard Cox using Boolean logic \cite{Cox1946}. The first three axioms are compatible with the usual probability rules. Here we have an additional axiom that, as we will see,  plays a very important role.

From these axioms the following rule (sum rule) is deduced \cite{Ballentine1986}
\begin{equation}
P(X + Y | I) = P(X | I) + P(Y | I) - P(X,Y | I). \label{eq:cox5}
\end{equation}
Here the symbol `$+$' 
in the notation $X + Y$
means here the logical disjunction ($X +Y \equiv X \vee Y \equiv$ ``X OR Y is true'').

The fourth axiom determines the rule for inference probabilities (product rule) for conditional cases. If $X$ and $Y$ are independent assertions, this is reduced to the classical probability property 
\begin{equation}
P(X,Y | I) = P(X | I) \times P(Y | I).
\end{equation}

When a set of mutual exclusive assertions are considered $\{Y_i\}$, with $P(Y_i | Y_{j \ne i}) = 0$, we have the so-called \textit{marginalization rule} 
\begin{equation}
P(X | I) = \sum_i P(X,Y_i | I)
\end{equation}
that in the limit of continuous case $Y_{i+1}-Y_i \to dY$ becomes
\begin{equation}
P(X | I) =  \int_{-\infty}^\infty P(X,Y | I) d Y. \label{eq:marg}
\end{equation}

\subsection{Bayes' theorem and  posterior probability}
Another important corollary can be derived from the fourth axiom (Eq.~\eqref{eq:cox4}) and the similar expression with exchange between $X$ and $Y$:
\begin{equation}
P(X | Y, I) = \frac{P(Y | X, I) \times P(X | I)}{P(Y | I)}.
\label{eq:bayes}
\end{equation}
This is what is called the Bayes' Theorem, named after Rev. Thomas Bayes, who first \cite{Bayes1763} formulated theorems of conditional probability, and rediscovered in 1774 and further developed by Pierre-Simon Marquis de la Laplace \cite{Laplace1825}.

For a better insight in the implication of this theorem, we consider the case where $X$ represent the hypothesis that the parameter values set $\boldsymbol{a}$ truly describes the data (via the function $F(x,\boldsymbol{a})$) and where $Y$ correspond to the recorded data $\{x_i,y_i\}$.
In this case Eq.~\eqref{eq:bayes} becomes
\begin{equation}
P(\boldsymbol{a} | \{x_i, y_i\},I) = \frac{P(\{x_i, y_i\} | \boldsymbol{a}, I) \times P(\boldsymbol{a} | I)}{P(\{x_i, y_i\} | I)} = 
\frac{L(\boldsymbol{a}) \times P(\boldsymbol{a} | I)}{P(\{x_i, y_i\} | I)}, \label{eq:bayes-exp}
\end{equation}
where $I$ includes our available background information and where $P(\{x_i, y_i\} | \boldsymbol{a}, I)$ is by definition the likelihood function $L(\boldsymbol{a})$ for the given set of data.
Differently from the common statistical approach where only the likelihood function is considered, we have here the additional term $P(\boldsymbol{a} | I)$ that includes the prior knowledge on the parameters $\boldsymbol{a}$ or its possible boundaries. 
The denominator term $P(\{x_i, y_i\} | I)$ can be considered for the moment as a normalization factor but it plays an important role when different hypotheses are considered and compared (see next section).

The priors can look as an unsuitable input due to the possible subjectivity in their choice; this is actually the main critics to \textit{Bayesian statistics}. On the contrary, the priors reflects our knowledge or ignorance in a quantify way.  If two scientists have different choices of priors, and uses some common experimental data, the posterior probability distributions are generally not significantly different.  If the posteriors are different because of the different choice of priors, this means that the data are not sufficient to analyze the problem.

From $P(\boldsymbol{a} | \{x_i, y_i\},I)$, the probability distribution of each parameter $P(a_j|\{x_i, y_i\},I)$ or joint probabilities $P(a_j,a_k|\{x_i, y_i\},I)$ can be obtained from the marginalization  (Eq.~\eqref{eq:marg}), i.e. the integration of the posterior probability on the unconcerned parameters.

For a more in-deep introduction to Bayesian statistics, we invite the reader to consult Refs.~\citenum{Jeffreys,Trotta2008,Sivia}.
In the following paragraphs, we will present more specific examples adapted to cases commonly encountered in atomic physics and related to the problem of hypotheses testing.

\subsection{Model testing and Bayesian evidence} \label{sec:model_testing}

An important consequence of the Bayes' theorem is to have the possibility to assign probabilities to different hypotheses (models) with a simple and well-defined procedure.
In this case, $X$ in Eq.~\eqref{eq:bayes}  represents the hypothesis that the model $\mathcal{M}$ describes well the observations and $Y$ represents the data, as in the previous section. 
From Bayes' theorem we have that the posterior probability of the model $\mathcal{M}$ is \cite{Jaynes,Sivia,Trotta2008}
\begin{equation}
P(\mathcal{M} | \{x_i, y_i\} ,I) \propto P( \{x_i, y_i\} | \mathcal{M}, I) \times P(\mathcal{M} | I), \label{eq:bayes-data}
\end{equation}
where the first term of the right part is the so-called \textit{Bayesian evidence} $E$ of the model and the second term is the prior probability assigned to the model from our background knowledge.
Using the marginalization rule to the parameter values and the probability properties (Eqs.~(\ref{eq:cox1}--\ref{eq:cox4})), we have
\begin{multline}
E \equiv P( \{x_i, y_i\} | \mathcal{M}, I) = \\
=  \int P( \{x_i, y_i\} | \boldsymbol{a},\mathcal{M}, I) P(\boldsymbol{a}| \mathcal{M}, I) d^{J}\boldsymbol{a} = \\
  = \int L^\mathcal{M}(\boldsymbol{a}) P(\boldsymbol{a}| \mathcal{M}, I) d^{J}\boldsymbol{a}, \label{eq:evidence}
\end{multline}
where $J$ is the number of the parameters of the considered model, and where we  explicitly show the likelihood function $L^\mathcal{M}(\boldsymbol{a})$ relative to the model $\mathcal{M}$. 
The Bayesian evidence, also called \textit{marginal likelihood} or  \textit{model likelihood}, is the the integral of the likelihood function over the $J$-dimensional parameter space under the priors constraints for a specific model choice.
The evidence is also the denominator of Eq.~\eqref{eq:bayes}, which now assumes a clearer signification than a simple normalization factor (with $\mathcal{M}$ included in $I$).
Considering equal priors, the probability of a model is higher if the evidence is higher, which means that the average of the likelihood function over the model parameter space is higher. 
To note, this does not implies that the maximum of the likelihood function is larger, as in the case of the likelihood ratio test used to compare the goodness of fit of two models (where however we do not assign probabilities to the models themselves but where we define only a criterion to choose between two models).
Models with higher number of parameters are generally penalized because of the higher dimensionality of the integral that corresponds to a larger parameter volume $V_{\boldsymbol{a}}$ (and then to a lower average value of the likelihood function).
In fact, the calculation of the model probability via the Bayesian evidence includes, in some sense, the Ockham's razor\footnote{\textit{``Non sunt multiplicanda entia sine necessitate''} , "Entities must not be multiplied beyond necessity" from William of Ockham's (1287-1347), which can be interpreted in a more modern form as ``Among competing hypotheses, the one with the fewest assumptions should be selected''.} favoring simpler models when the values of the likelihood function are similar.

If we have to choose among only two different models $\mathcal{M}_1,\mathcal{M}_2$, the comparison between model probabilities is related to the calculation of the simple ratio
\begin{equation}
\frac{P(\mathcal{M}_1 | \{x_i, y_i\} ,I)}{P(\mathcal{M}_2 |\{x_i, y_i\} , I)} = \frac {P( \{x_i, y_i\} | \mathcal{M}_1, I)}{P( \{x_i, y_i\} | \mathcal{M}_2, I)} \times \frac{P(\mathcal{M}_1 | I)}{P(\mathcal{M}_2 | I)}.
\end{equation}
If the prior probabilities of the models are equal, this probability ratio is given by the \textit{Bayes factor} $B_{12}=E_1/E_2$ that is nothing else than the ratio of the evidences \cite{Jeffreys,Sivia,Trotta2008}.
Values of $B_{12}$ larger or smaller than one indicate a propensity for $\mathcal{M}_1$ or $\mathcal{M}_2$, respectively.
In the literature several tables are available to assign, in addition to probabilities, degrees of propensity of favor to one or other model \cite{Jeffreys,Kass1995} with a correspondence to the p-value and the standard deviation \cite{Gordon2007}.

For models with similar values of evidence, another criterium to decide between them is the Bayesian complexity $\mathcal{C}$, which measures the number of model parameters that the data can support \cite{Trotta2008}. 
This quantity is related to the gain of information (in the Shannon sense) and it is discussed in \ref{app:info}.
When $E$ values are similar, we should favor the simplest model, i.e. the model with the smallest $\mathcal{C}$.

The possibility to assigning probabilities to models has another important advantage.
In the case we are interested to determine the probability distribution of a common parameter $a_j$  without the need to identify the correct model among the available choices $\mathcal{M}_\ell$, we can obtain the probability distribution $P(a_j | \{x_i, y_i\},I)$ from the weighted sum
\begin{equation}
P(a_j | \{x_i, y_i\},I) = \sum_\ell P(a_j | \{x_i, y_i\},\mathcal{M}_\ell,I) \times P(\mathcal{M}_\ell |,I),  \label{eq:average}
\end{equation}
where $P(a_j | \{x_i, y_i\},\mathcal{M}_\ell,I)$ are the probability distributions of $a_j$ for each model and $P(\mathcal{M}_\ell |,I)$ are the probabilities of the different models.
As we will see in Sec.~\ref{sec:nasty_peak}, this capability plays an important role in the case where models have comparable probabilities.

\section{The nested sampling algorithm} \label{sec:nested}

\subsection{The evidence calculation problem}
The major difficulty to calculate hypothesis probabilities is the substantial computational power required for the evaluation of the Bayesian evidence.
Contrary to the maximum likelihood method, where only the maximum of a function has to be found, we have to calculate an integral over the $J$-dimensional space of parameters $V_{\boldsymbol{a}}$.
Except in very few cases, there is not analytical solution of Eq.~\eqref{eq:evidence}.
Numerical integration by quadrature is not efficient due to the span of different order of magnitude of the likelihood function and the high dimensionality of the problem.
The calculation of the evidence is then generally done via the Monte Carlo sampling of the product $P( \{x_i, y_i\} | \boldsymbol{a},\mathcal{M}, I) P(\boldsymbol{a}| \mathcal{M}, I)$.

A common approach to produce good sampling is the use of the Markov chain Monte Carlo (MCMC) technique.
A Markov chain is a sequence of random variables such that the probability of the $n^\text{th}$ element in the chain only depends on the value of the $(n-1)^\text{th}$ element.
The purpose of the Markov chain is to construct a sequence of points $\boldsymbol{a}_n$ in the parameter space whose density is proportional to the posterior probability distribution.
Different probabilistic algorithms are applied to build these chains like Metropolis-Hasting algorithm, Gibbs sampling, Hamiltonian Monte Carlo, etc. (see as example Ref.~\cite{Robert} and references their-in).

Another method is the  \textit{nested sampling}, originally developed by John Skilling in 2004 \cite{Skilling2004,Sivia,Skilling2006}.
On this method is based the program \texttt{Nested\_fit}, the Bayesian data analysis program we present in this article.
The method algorithm is based on the subdivision of the parameters space volume $V_{\boldsymbol{a}}$, delimited by the parameters prior probabilities, into $J$-dimensional nested volumes that get closer and closer to the maxima of the likelihood function. With this method, the calculation of the evidence (Eq.~\ref{eq:evidence}) is reduced to an one-dimensional integral from the original $J$-dimensional problem. 

\begin{figure}
\begin{center} 
\includegraphics[width=\columnwidth]{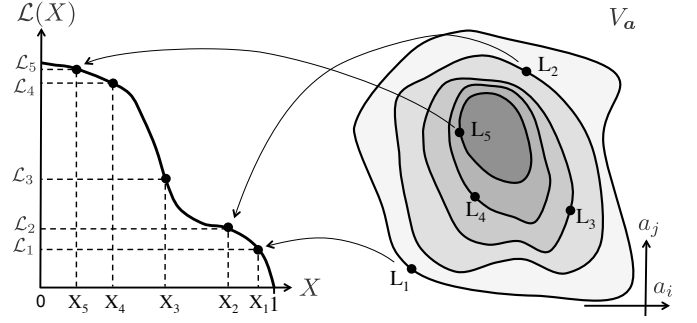} 
\caption{Visualisation of the integral of $\mathcal{L}(X)$  and corresponding volumes on the parameter space (two parameters only, $a_i,a_j$ are considered with a 2D representation).}
\label{fig:nested}
\end{center}
\end{figure}

To reduce to an one-dimensional integral, we define the variable $X$ (real and positive) corresponding to the volume of the parameter space, weighted by the priors, for which the likelihood function is larger than a certain value $\mathcal{L}$: 
\begin{equation}
X(\mathcal{L}) = \int_{L(\boldsymbol{a})>\mathcal{L}} P(\boldsymbol{a}| I) d^{J}\boldsymbol{a}. \label{eq:XvsL}
\end{equation}
A schematic visualisation of this relation is presented in Fig.~\ref{fig:nested}.
$X(\mathcal{L})$ is by construction monotonic and invertible, with $\mathcal{L} = \mathcal{L}(X)$.
When $\mathcal{L}=0$, the whole parameter volume $V_{\boldsymbol{a}}$ is considered and then $X=1$ because of the prior probability normalization.
When $\mathcal{L} \ge \max[L(\boldsymbol{a})]$, $X$ is equal to zero.
The infinitesimal volume $d X$ is 
\begin{equation}
d X  = P(\boldsymbol{a}| I) d^{J}\boldsymbol{a}, \label{eq:dXvsda}
\end{equation}
where $P(\boldsymbol{a}| I) d^{J}\boldsymbol{a}$ corresponds to the infinitesimal weighted volume of the parameter space where $\mathcal{L}(X)<L( \{x_i, y_i\}, \boldsymbol{a})<\mathcal{L}(X +d X)$.

With the above definitions, we can then rewrite Eq.~\eqref{eq:evidence} as a simpler one-dimensional integral in $X$:
\begin{equation}
E = \int _0^1 \mathcal{L}(X) d X. \label{eq:evX}
\end{equation} 

\subsection{The algorithm for the numerical integration}
The one-dimensional integral in the above equation and represented on the left part of Fig.~\ref{fig:nested} can be numerically calculated using the rectangle integration method subdividing the $[0,1]$ interval in $M+1$ segments with an ensemble $\{X_m\}$ of $M$ ordered points 
$0< X_M<...<X_2<X_1<X_0=1$. 
Equation~\eqref{eq:evX} is approximated by the sum
\begin{equation}
E \approx \sum_m \mathcal{L}_m \Delta X_m, \label{eq:evindence_app}
\end{equation} 
where $ \mathcal{L}_m = \mathcal{L}(X_m)$ and $\Delta X_m= X_m - X_{m+1}$.
The difficulty is now the determination of $\mathcal{L}_m$ and $\Delta X_m$ because we do not know a priori the relation between $X$ and $\mathcal{L}$. 

The evaluation of $\mathcal{L}_m$ values is obtained by the exploration of the likelihood function with a Monte Carlo sampling via a subsequence of steps.
For this, we use a collection of $K$ parameter values $\{\boldsymbol{a}_k\}$ that we call \textit{live points}. 
At the beginning, these values are randomly chosen from the prior probability distribution $P(\boldsymbol{a}_k| I)$  and they evolve during the computation steps described in the following paragraphs.

\begin{figure}
\begin{center} 
\includegraphics[height=7cm]{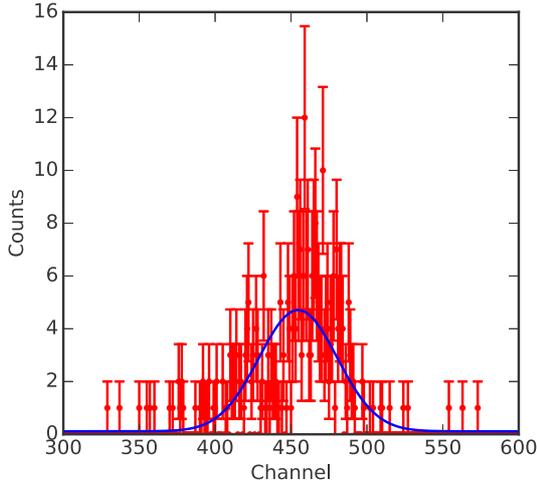} 
\caption{High-resolution X-ray spectrum of the helium-like uranium  $1s2p\ ^3\!P_2 \to 1s2s\ ^3\!S_1$ intrashell transition from Ref.~\cite{Trassinelli2009b} and corresponding fit with one Gaussian peak (plus a flat background).}
\label{fig:peak}
\end{center}
\end{figure}

To clearly present the different stages of the algorithm, we consider a real analysis of a very simple case.
We assume a Gaussian peak plus a flat background (four parameters in total) as model and a very statistical poor data set. The data refer to a high-resolution X-ray spectrum of the helium-like uranium $1s2p\ ^3\!P_2 \to 1s2s\ ^3\!S_1$  intrashell transition obtained by Bragg diffraction from a curved crystal \cite{Trassinelli2009b}.
The data and the best guess (maximum likelihood) of Gaussian peak profile are shown in Fig.~\ref{fig:peak}.

For each computation step $m$ of the algorithm we indicate with  $\{\boldsymbol{a}_{m,k}\}$ the live points of the step, with $k=1,\dots, K$.
The corresponding likelihood function values are indicated by $\mathcal{L}_{m,k} = L(\boldsymbol{a}_{m,k})$ and we define  $\mathcal{L}_m = \min(\mathcal{L}_{m,k})$.
The related $X$ values are indicated by $\xi_{m,k}=X(\mathcal{L}_{m,k})$ and we define $X_m = \max(\xi_{m,k})$.
Considering Eq.~\eqref{eq:XvsL} and Fig.~\ref{fig:nested}, we see that $X_m$ is equal to the integral of the volume where all $\{\boldsymbol{a}_{m,k}\}$ are contained.
In other words, the volume $V_{L \ge \mathcal{L}_m}$ in the parameter space corresponds to the segment $[0,X_m]$  in the $X$ axis.
Let us see the different steps of the algorithm in details.

\hfill

\textbf{Step 1:}
The initial $\{\boldsymbol{a}_{1,k}\}$ live points  are sorted considering $P(\boldsymbol{a}_k| I)$ and $\mathcal{L}_{1}=\min(\mathcal{L}_{1,k})$ is found.
From $\xi = X(\mathcal{L})$ relationship, we have $X_1 = \max(\xi_{1,k})$ and the  $\Delta X_1 = X_0 - X_1$, where $X_0=1$.
We have now our first pair of values for the sum in Eq.~\eqref{eq:evindence_app}.

\hfill

\textbf{Step 2:}
We built now a new ensemble of live points $\{\boldsymbol{a}_{2,k}\}$, which is the same as $\{\boldsymbol{a}_{1,k}\}$ but where we remove the $k'$-th element with the lower value of likelihood (corresponding to the higher value of $X$, i.e. where $\mathcal{L}_1= L( \boldsymbol{a}_{1,k'})$ with $X_1 =  \xi_{1,k'}$, and we store its value with the name $\boldsymbol{\tilde a}_1 = \boldsymbol{a}_{1,k'}$.
We replace this point with a new $\boldsymbol{a}$ value, randomly chosen with the only condition $L(\boldsymbol{a})>\mathcal{L}_1$.
With this requirement we impose that this point is inside the volume $V_{L \ge \mathcal{L}_1}$.
From this new ensembles $\{\xi_{2,k}\}$ and $\{\mathcal{L}_{2,k} \}$ we define $X_{2}= \max (\xi_{2,k})$.
The interval $[0,X_2]$ corresponds to the volume of the parameter space $V_{L \ge \mathcal{L}_2}$ nested in the volume $V_{L \ge \mathcal{L}_1}$ (see Fig.~\ref{fig:nested}).
We have then the elements $\mathcal{L}_2, \Delta X_2$ of the sum in Eq.~\eqref{eq:evindence_app} and we store the value of the discarded live point $\boldsymbol{\tilde a}_1$.

\hfill

\textbf{Step m:}
We continue the iteration as in the step 2, storing at each step the values $\mathcal{L}_m,\Delta X_m$ and $\boldsymbol{\tilde a}_m$.
All new live points $\{\boldsymbol{a}_{m,k}\}$ are enclosed in smaller and smaller parameter volumes defined by $L(\boldsymbol{a})>\mathcal{L}_m$ that correspond to $[0,X_m]$ intervals (see Fig.~\ref{fig:nested}) with $X_m = \max (\xi_{m,k})$.

\hfill

\textbf{Step M, the end:}
After $M$ iterations, the estimated error $Err_{M}$ on the evidence $E$ evaluation due to the truncation of the sum in Eq.~\eqref{eq:evindence_app} is less than the target accuracy $\Delta E$ and the calculation stops.
For each step $m$, $Err_{m}$ is upper limited by the product $L_{max} X_m$ where  $L_{max} = \max [ L(\boldsymbol{a}_{m,k}) ]$. 
When $L_{max} X_m < \Delta E$, we have $Err_{m} < \Delta E$, the main iteration loop of the nested algorithm stops and the main calculation is finalized.
The likelihood function value associated to the last live points is the average $\mathcal{L}_M = \langle L(\boldsymbol{a}_{M,k}) \rangle$.

\hfill

\begin{figure}
\begin{center} 
\includegraphics[height=7cm]{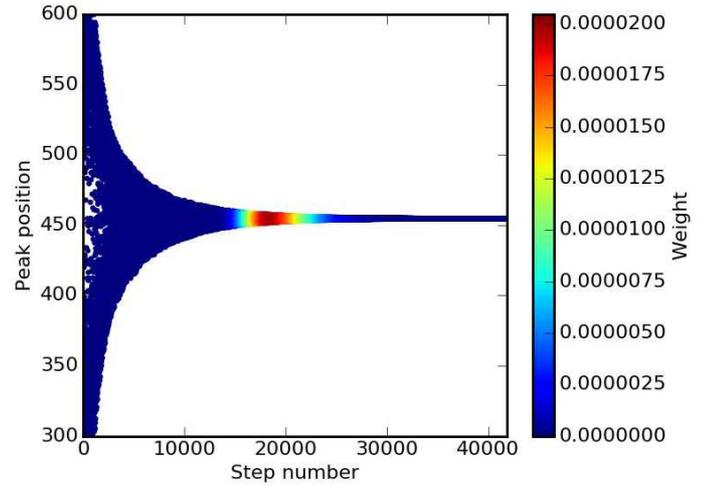} 
\caption{Evolution of the sampled parameter value relative to the peak over the algorithm step.}
\label{fig:step_pos}
\end{center}
\end{figure}

In addition to the final live points $\{\boldsymbol{a}_{M,k}\}$ and their likelihood function values, all intermediate $\mathcal{L}_m,\Delta X_m, \boldsymbol{\tilde a}_{m}$ are stored and used for the calculation of the posterior probability distributions as presented in Sec.~\ref{sec:posterior}.

\begin{figure}
\begin{center}
\includegraphics[height=7cm]{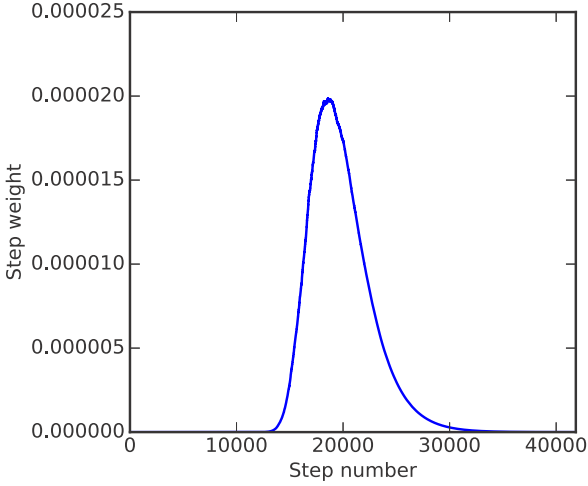} 
\caption{Weights associated to discarded value at each step, which are proportional to the product $\mathcal{L}_m  \Delta X_m)]$.
}
\label{fig:step_weights}
\end{center}
\end{figure}

For the specific example where we consider the analysis of the data in Fig~\ref{fig:peak} and a Gaussian peak as model, we show in Fig.~\ref{fig:step_pos} the evolution of the values of the $\boldsymbol{\tilde a}_{m}$, corresponding in this example to the peak position, as function of the algorithm step number.
Starting from a sampling range corresponding to our priors (here a flat distribution between channel 300 and 600), the algorithm explores smaller and smaller ranges corresponding to nested volumes of the parameter space.
The product $\mathcal{L}_m \Delta X_m$ relative to each steps are shown in both plots of Fig.~\ref{fig:step_weights} via the value  $\textit{weight} =\mathcal{L}_m  \Delta X_m / E$ (see next section for further explanation).

We have now a recipe for calculating $\mathcal{L}_m$ values but not the $X_m$. 
In the previous paragraphs we defined $X_m = \max[X(\mathcal{L}_{m,k})]$ using Eq.~\eqref{eq:XvsL}.
But we do not know the function $X(\mathcal{L})$ and neither its inverse $\mathcal{L}(X)$.
We can, however, estimate the values of $X_m$ from some statistical consideration.
The extraction of a set of $K$ live points $\boldsymbol{a}_{m,k}$ in the parameter volume $V_{L(\boldsymbol{a})>\mathcal{L}_m}$ correspond to sort $K$ random numbers in the interval $[0,X_m]$ (with $\xi_{m,k} = X[L(\boldsymbol{a}_{m,k})]$). 
For each step, when we pass from the $[0,X_{m-1}]$ interval to the $[0,X_{m}]$ interval, we  shrink the volume (one-dimensional here) by a factor $t_m=X_m/X_{m-1}$.
The probability distribution for each $t_m$ is equal to the probability for having a maximum value $t$ given $K$ random numbers $\in [0,1]$.
The statistical distribution of $t$ is (see \ref{app:uncertainty} for more details) 
\begin{equation}
P(t)=K t^{K-1}, \quad \text{ with } \quad \langle \ln t \rangle = -1 / K. \label{eq:Pt}
\end{equation}
For the first and second step we have $X_1 = t_1$ ($X_0=1$) and $X_2 = t_2 X_1 = t_1 t_2 $. 
For a  generic step, considering Eq.~\eqref{eq:Pt}, $X_m$ is given by the product 
\begin{equation}
X_m = \prod_i^m t_i  \quad \text{ and then } \quad X_m \approx e^{-m/K} \label{eq:Xm}.
\end{equation}
From this equation, the values of $\Delta X_m$ can be estimated, with $\Delta X_M = e^{-M/K}$ for the last live points.
This approximation introduces an error in the evidence calculation that is proportional to $1/\sqrt{K}$, where $K$ is the number of the employed live points.
A detailed discussion of the evidence uncertainty is presented in \ref{app:uncertainty}.

We note that for the final calculation of the evidence, the terms $\mathcal{L}_m, \Delta X_m$ in Eq.~\eqref{eq:evindence_app} are not equally important.
$\Delta X_m$ values are monotonically decreasing with $m$ where $\mathcal{L}_m$ values are increasing. 
As we can see from Fig.~\ref{fig:step_pos}, the product $\mathcal{L}_m  \Delta X_m$ (which defines the step weight as we will see next section) has a maximum.
$\boldsymbol{\tilde a}_{m}$ corresponding to this maximum will strongly influence the posterior probability distributions and the value of the evidence.

\hfill

The bottleneck of the nested sampling algorithm is the search of new points within the $J$-dimensional volume defined by $L>\mathcal{L}_m$.
Different methods are commonly employed to accomplish this difficult task. 
One efficient method is the ellipsoidal nested sampling \cite{Mukherjee2006}. It is based  for each step on the approximation of the iso-likelihood contour defined by $L =\mathcal{L}_m$ by a $J$-dimensional ellipsoid calculated from the covariance matrix of the live points.
The new point is then selected within the ellipsoidal volume (times an user-defined enlargement factor).
This methods, well adapted for unimodal posterior distribution has been also extended to multimodal problems \cite{Feroz2008,Feroz2009}, i.e. with the presence of distinguished regions of the parameter space with high values of the likelihood function. 
Other search algorithms are based on Markov chain Monte Carlo (MCMC) methods \cite{Veitch2010},  as in particular the \textit{lawn mower robot} method, developed by L.~Simons \cite{Theisen2013}, and the recent 
\textit{Galilean Monte Carlo} \cite{Skilling2012,Feroz2013}, particularly adapted to explore the regions close to the boundary of $V_{L>\mathcal{L}_m}$ volumes.
\texttt{Nested\_fit} program is based in an evolution of Simons' algorithm and is presented in Sec.~\ref{sec:nested_fit}.

Additional material on the nested sampling can be found in Refs.~\citenum{Skilling2004,Sivia,Skilling2006,Mukherjee2006,Feroz2009,Veitch2010}.
In particular in Ref.~\citenum{Buchner2016}, the different search algorithms, their efficiency and accuracy are discussed.

\begin{figure}
\begin{center} 
\includegraphics[height=7cm]{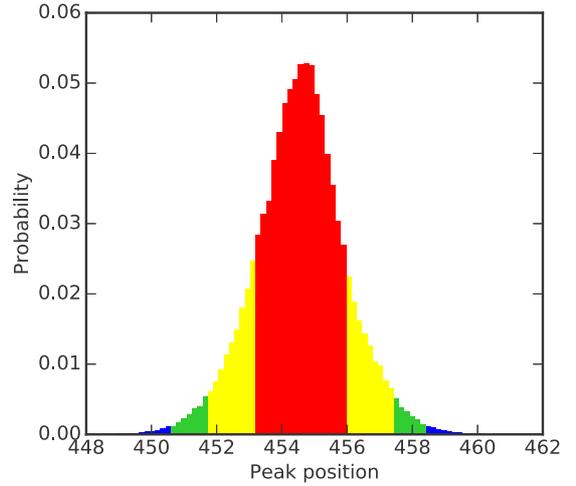} 
\caption{Histogram of the Gaussian peak position built from the values $\mathcal{L}_m,\Delta X_m$, and $\boldsymbol{\tilde a}_{m}$ (see text).
Red, yellow and green regions indicate 68\%, 95\% and 99\% confidence intervals (credible intervals).}
\label{fig:histo}
\end{center}
\end{figure}

\subsection{Posterior probability distributions} \label{sec:posterior}
The posterior probability distributions are built from the discarded live points $\boldsymbol{\tilde a}_{m}$, the final set of $K$ live points $\boldsymbol{a}_{M,k}$ and their associated $\mathcal{L}_m, \Delta X_m$ values.

Once the evidence $E \equiv P( \{x_i, y_i\} | I)$ is determined, posterior inference can be easily generated from the $\{\boldsymbol{\tilde a}_{m}\}$ and $\{\boldsymbol{a}_k\}_M$ values.
Each $\boldsymbol{\tilde a}_{m}$ is in the infinitesimal parameter volume $\Delta V_{\mathcal{L}_m <L(\boldsymbol{\tilde a}_{m})<\mathcal{L}_{m+1}}$ that correspond to the interval $\Delta X_m$. 
Considering the discrete form of Eq.~\eqref{eq:dXvsda} and Eq.~\eqref{eq:bayes-exp}, we can calculate the probability associated to the parameter values $\boldsymbol{\tilde a}_{m}$, in other words the step \textit{weight} named in the previous sections:
\begin{equation}
P(\boldsymbol{\tilde a}_{m} | \{x_i, y_i\},I) = P(X_m) \approx \frac{\mathcal{L}_m \Delta X_m} E. \label{eq:Pa}
\end{equation}
From Eq.~\eqref{eq:Pa}, the probability distribution of any single parameter $a_j$ is obtained by marginalization (Eq.\eqref{eq:marg}), 
i. e. integrating of the posterior probability $P(\boldsymbol{a} | \{x_i, y_i\},I)$ over the other parameters.
In our case, if the parameter of interest corresponds to the $j^\text{th}$ component, its probability distribution can be built from $ (\tilde a_m)_j$ values and their corresponding weights defined by Eq.~\eqref{eq:Pa}.

\begin{figure}
\begin{center} 
\includegraphics[height=7cm]{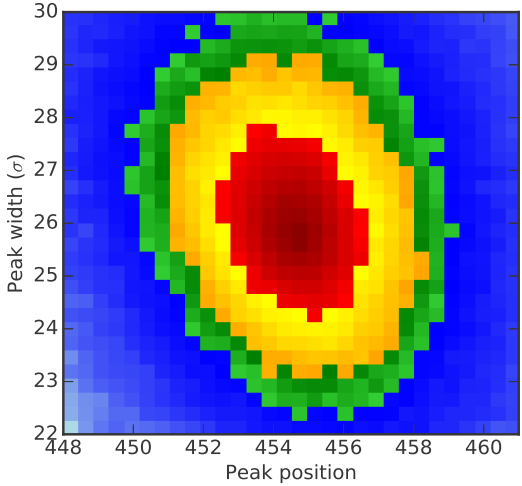} 
\caption{Joint probability distribution of the parameters relative to the peak position and width (in terms of sigmas of the Gaussian profile).
Red, yellow and green regions indicate 68\%, 95\% and 99\% confidence intervals (credible intervals).}
\label{fig:histo2D}
\end{center}
\end{figure}

For our specific example with a Gaussian distribution as a model, this corresponds to take $(\tilde a_m)_j$ values  showed in Fig.~\ref{fig:step_pos} (top) and built a weighted histogram (with the weights $\mathcal{L}_m \Delta X_m / E$ showed by the different color intensities in Fig.~\ref{fig:step_pos} and in Fig.~\ref{fig:step_weights}).  
From the marginalization on $J-2$ parameters, also joint probabilities can be built, as that one presented in Fig.~\ref{fig:histo2D} corresponding to the position and width distribution of the peak.

\section{The \texttt{Nested\_fit} program} \label{sec:nested_fit}

\subsection{General considerations}

\texttt{Nested\_Fit} has been developed in Fortran90 for the calculation of the Bayesian evidence and posterior parameters probability distributions for a given set of data and selected model. 
The core of  \texttt{Nested\_Fit} is the algorithm used for the calculation of the Bayesian evidence which is, as indicated by its name, the nested sampling developed by Skilling and presented in Sec.~\ref{sec:nested}, but with an original method to find new live points.

Due to the large span of the values of the different quantities (likelihood function, evidence, $X_m$, etc.), all computations are done with respect to their logarithms, as many programs based on the nested sampling.
The data inputs are provided in the form $\{x_i,y_i\}$, where $x_i$ are real numbers and $y_i$ are necessarily counts detected at the channels $x_i$.  
To analyze statistically poor (but also not-poor) data sets, the likelihood function is built considering a Poissonian statistics for each channel (which tend to the normal distribution for large number of counts), leading to
\begin{equation}
L(\boldsymbol{a}) = \prod_i \frac{ F(x_i,\boldsymbol{a})^{y_i} e^{-F(x_i,\boldsymbol{a})}} {y_i !},
\end{equation}
where for each channel, $y_i$ is the recorded number of counts and $F(x_i,\boldsymbol{a})$ is the expected value of the modeling function that depends on the parameters $\boldsymbol{a}$.
A large library of functions is available and new \textit{ad hoc} functions can easily be added.

Outputs of \texttt{Nested\_Fit} include the evaluation of the Bayesian evidence, the corresponding information gain and complexity, and the information to build parameter probability distributions. 
The different probability histograms and other plots are produced via a series of functions of a dedicated Python library. 
The figures of this article are examples of their typical outcomes.
In addition to the graphic outputs, Python library functions can be used to recursively modify the input file \texttt{nf\_input.dat} and to read the results in the output files. 
These functions are particularly useful for automated analysis and systematic surveys.

Several set of data can be analyzed at the same time by \texttt{Nested\_Fit} program. 
For example, distinct spectra with a same response function can be analyzed, and common parameters such as the profile width can be extracted by correctly taking into account the correlations between data sets.

\subsection{Computation algorithm of the Bayesian evidence}
The calculation of the Bayesian evidence is made with the nested sampling following the steps presented in Sec.~\ref{sec:nested}, similarly to other programs based to the same algorithm \cite{Sivia, Mukherjee2006,Feroz2008,Feroz2009,Veitch2010}.
Even if the basic structure is practically identical to existing codes, the algorithm for the search of new live points is substantially different. 
The searching algorithm is a Markov chain Monte Carlo method to explore the parameter volume $V_{L>\mathcal{L}_m}$ and it is an evolution of the \textit{lawn mower robot} method, developed by L.~Simons \cite{Theisen2013}.
To cancel the correlation between the starting point and the final point, a series of $N$ jumps are done in this volume.
The different stages of the algorithm are 
\begin{enumerate}
\item Choose randomly a starting point $\boldsymbol{a}_{n=0} = \boldsymbol{a}_0$ from the available live points $\{\boldsymbol{a}_{m,k}\}$ as starting point of the Markov chain where $n$ is the number of the jump. 
The number of tries $n_t$ (see below) is set to zero.

\item From  the values $\boldsymbol{a}_{n-1}$, find a new parameter sets $\boldsymbol{a}_{n}$  where each $j^\text{th}$ parameter is calculated by $(a_{n})_j= (a_{n-1})_j +  f\ r\ \sigma_j$, where $\sigma_j$ is the standard deviation of the live points $\{\boldsymbol{a}_{m,k}\}$ relative to the $j^\text{th}$ parameter, $r \in [-1,1]$ is a sorted random number and $f$ is a factor defined by the user.

\begin{enumerate}
\item If $L(\boldsymbol{a}_{n})>\mathcal{L}_m$ and $n< N$, go to the beginning of step 2 with an increment of the jump number $n = n + 1$.

\item If $L(\boldsymbol{a}_{n})>\mathcal{L}_m$ and $n= N$, $\boldsymbol{a}_{n=N}$ is new \textit{live point} to be included in the new set $\{\boldsymbol{a}_{m+1,k}\}$.

\item If $L(\boldsymbol{a}_{n})<\mathcal{L}_m$ and $n< N$ and the number of tries $n_t$ is less than the maximum allowed number $N_t$, go back to beginning of step 2 with an increment of the number of tries $n_t = n_t + 1$.

\item If $L(\boldsymbol{a}_{n})<\mathcal{L}_m$ and $n< N$ and $n_t = N_t$ a new parameter set $\boldsymbol{a}_0$ has to be selected.
Instead than choosing one of the existing live points, $\boldsymbol{a}_0$ is built from distinct $j^\text{th}$ components from different live points: $(a_0)_j = (a_{m,k})_j$ where $k$ is randomly chosen between 1 and $K$ for each $j$. 
 Then $\boldsymbol{a}_{n=0} = \boldsymbol{a}_0$ and go to the beginning of step 2.

\end{enumerate}

\end{enumerate}

\begin{figure}
\begin{center} 
\includegraphics[height=8cm]{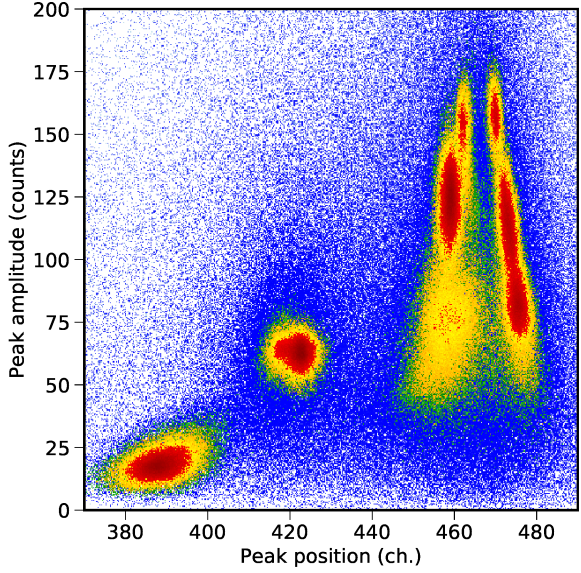} 
\caption{Joint probability distribution of the parameters relative to the position and amplitude of one peak in four Gaussian peaks model. Eight distinct likelihood maxima can be identified.
Red, yellow and green regions indicate 68\%, 95\% and 99\% confidence intervals.}
\label{fig:histo_peaks}
\end{center}
\end{figure}

The last step makes the algorithm well adapted to problems with multimodal parameter distributions allowing to easily jump between high-likelihood regions. 
An example of presence of several maximal likelihood regions is presented in Fig.~\ref{fig:histo_peaks} where we plot the joint probability of the position and amplitude of one of the four Gaussian peaks of the considered model. 
The value of $N_t$ is fixed in the code ($N_t = 10000$ in the present version). 
The other parameters can be provided by the input file.

\subsection{Inputs}
All input parameter required by \texttt{Nested\_fit} are provided in the file \texttt{nf\_input.dat}.
The most important inputs are:

\begin{description}

\item[The maximum number of jumps $\boldsymbol{N}$ and the real number $\boldsymbol{f}$:]
These parameters are important for efficiency of the search of the new live points and for the non-correlated and efficient exploration of the parameter space. A higher value of $f$ guarantees a better independence between the current live points and the new point but also a minor efficiency for finding it because of the higher probability to jump in the volume region $L<\mathcal{L}_m$. The same reasoning applies for the total number of jumps $N$.

\item[The number of live points $\boldsymbol{K}$:]
The choice of $K$ influence directly the expected accuracy of the evidence $\delta E \propto 1/\sqrt{K}$, and also provides a better sampling of the parameter volume.
As counterpart, an increasing of $K$ increases the computation time.

\item[The required final evidence accuracy $\boldsymbol{\Delta E}$:] A too large value of the accuracy will bias the evidence calculation. A too small value can make the evidence computation significantly long. For a given problem, the optimal value is obtained by looking a posteriori at the evolution of $\mathcal{L}_m \Delta X_m$. The calculation has  to stop significantly far from the region where the product $\mathcal{L}_m \Delta X_m$ is large, i. e. far from the most influent values of $X \sim \exp(-\mathcal{H})$ where $\mathcal{H}$ is the extracted information (in the sense of Shannon, see \ref{app:info} and \ref{app:uncertainty}). Good and efficient values are generally between $10^{-3}$ and $10^{-5}$ as also discussed in Ref.~\citenum{Veitch2010}.

\item[The number of trials sets of live points $\boldsymbol{N_{LPS}}$:] Besides theoretical considerations, the best strategy to estimate the evidence accuracy is to calculate $E$ several times with different starting sets of live points (with different seed for the random generator) and to extract the mean and standard deviation of the logarithmic values of the computed evidence, which is the pertinent quantity for the uncertainty evaluation (see \ref{app:uncertainty}). 
In addition this method provides more sampling points of the parameter space for a better evaluation of the posterior probability distributions, especially important when multimodal distributions are present.

\item[The parameter priors] Priors of the different parameters can be selected between two options: (i) an uniform prior where the parameter value boundaries have to be provided or (ii) a normal distribution where a main value and the associated standard deviation have to be provided (as example from a past experiment).

\end{description}

Except for the priors, each parameter has to be tuned by looking the output in order to have valuable results (to uniformly and randomly cover the entire parameter space) but also to have a fast calculation (a good efficiency to find new live points).
For this goal, the most sensitive parameters are the number of live points $K$, the number of jumps $N$ and the real number $f$.

\subsection{Outputs}
Once ended, the program provides four major output files described below.
\begin{itemize}

\item \texttt{nf\_output\_res.dat} contains the details of the computation (n. of live points trials, n. of total iteration), the final evidence value and its uncertainty $E \pm \delta E$, the parameter values $\boldsymbol{\hat a}$ corresponding to the maximum of the likelihood function, the mean, the median, the standard deviation and the confidence intervals (68\%, 95\% and 99\%) of the posterior probability distribution of each parameter.
Moreover, the information gain $\mathcal{H}$, the Bayesian complexity $\mathcal{C}$ and the theoretical minimal value of the required iteration number deduced from the computed information gain $\mathcal{H}$, also provided in the output.
$\delta E$ is calculated only if $N_{LPS}>2$.

\item \texttt{nf\_output\_data.dat} contains the original input data together with the model function values corresponding to the parameters $\boldsymbol{\hat a}$ with the highest likelihood function value, the residuals and the uncertainty associated to the data.

\item \texttt{nf\_output\_tries.dat} is present only if $N_{LPS}>2$. For each live points trial, it contains the final evidence, the number of iterations and the maximum value of the likelihood function.

\item \texttt{nf\_output\_points.dat} contains all discarded and final live points values $\boldsymbol{\tilde a}_m$ and $\{\boldsymbol{a}_{M,k}\}$, their associated  likelihood values $L( \boldsymbol{a})$ and posterior probabilities $P(\boldsymbol{a}| \{x_i, y_i\},I) \approx \mathcal{L}_m \Delta X_m / E$.
From them, the different parameter probability distributions, as shown in Fig.~\ref{fig:histo}, or joint probabilities, as shown in Figs.~\ref{fig:histo2D} and \ref{fig:histo_peaks}, can be built from the marginalization (Eq.~\ref{eq:marg}) of the unretained parameters.

\end{itemize}

\section{Two examples}\label{sec:applications}
In this section we will present two practical applications of the of the statistical analysis methods described above.
In the first one, we calculate the probability of the presence or not of a satellite line in a spectrum at a well defined position but with unknown intensity.
The second, more complex, consists in the analysis of a statistically poor set of data for which we would like to determine the most probable model among different possibilities and to extract the position of the main component.

\subsection{Satellite line contamination}

\begin{figure}
\begin{center} 
\includegraphics[height=7cm]{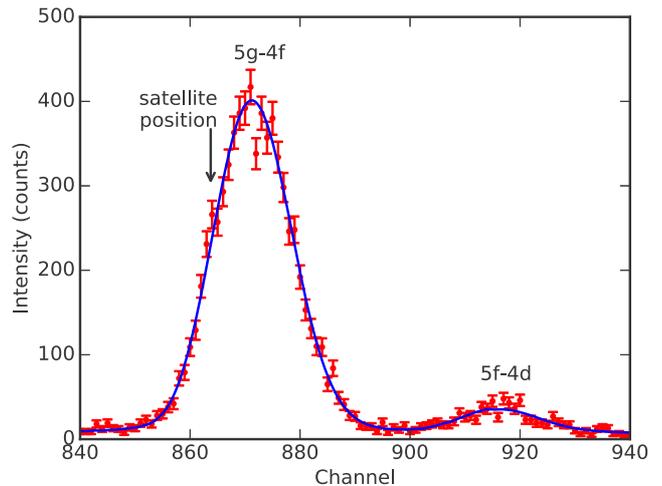} 
\caption{Pionic nitrogen $5-4$ transitions. Possible additional transitions from the presence of one remaining electron in the K shell are indicated}
\label{fig:piN}
\end{center}
\end{figure}

We consider a common case in spectroscopy where we would like to test the presence or not of an unresolved weak spectral line close to a intense line.
In this specific example, we consider the $5g-4f$ transition in pionic nitrogen, an  hydrogen-like atom formed by a nitrogen nucleus and a negatively charged pion.
During the formation of the pionic atoms, all electrons are expected to be ejected. 
The presence of a remaining electron in the $K$ shell cannot completely be excluded. Its presence can cause a shift of the main tradition energy due to the Coulomb screening and then an appearance of a new component in the spectrum.
To determine the probability of such a scenario, we have to calculate the evidence for the two possible models: Model 1 without remaining electrons (a pure hydrogen-like pionic atom) and Model 2 with the possible presence of one remaining electron.
More details on the physics case can be found in Refs. \citenum{Trassinelli2016b,Trassinelli2016c}.

The examined data consist in seven distinct spectra similar to the one represented in Fig.~\ref{fig:piN} with a total of about 60000 recorded counts .
Each spectra is obtained by a Bragg spectrometer equipped by a spherically bent crystal.
The evidence and probability distributions of both models are computed with \texttt{Nested\_fit} taking into account all seven spectra at the same time.
For this specific propose we used $K = 1000$ live points and an accuracy requirement $\delta E = 10^{-5}$. 
For the search of the new points we choose the values $J=20$ jumps and $f=0.1$. These parameters insure an efficient and complete exploration of the parameter space and an accurate evaluation of the evidence.
For a rough estimation of the evidence uncertainty we consider $N_{LPS} = 8$ different live point trial sets.
For both models, we chose flat prior probability distributions for the different parameters. Compared to model 1, model 2 has only as additional free parameter the satellite line intensity whose relative position with respect to the main line has been fixed by the theory.

Since we have to choose among two models only, the relevant quantity to calculate is the Bayes factor $B_{12}$, defined in Sec.~\ref{sec:model_testing}, from which we can determine the criterium in favor to one of the two hypothesis.

From the output of \texttt{Nested\_fit}, we obtain $\ln B_{12} = 6.6 \pm 1.8$, which correspond to a probability of 99.98\% in favor to the model without remaining electrons (between 99.86\%  and 100\% when the Bayes factor uncertainty is taken into account). 
This Bayes factor value indicates a decisive support for the Model 1 hypothesis considering any considered scale (``decisive'' in the Jeffreys scale \cite{Jeffreys}, ``very strong''  in the Kass scale \cite{Kass1995} or ``strong'' in the Gordon-Trotta scale \cite{Gordon2007}) with an equivalent p-value of about $10^{-5}$ for Model~2 \cite{Gordon2007}.


In conclusion, the presence of remaining electrons can be safely excluded and the main line position can be reliably evaluated.
Additional discussion on this analysis can be found in Ref.~\citenum{Trassinelli2016c}.

\subsection{A nasty peak} \label{sec:nasty_peak}

\begin{figure}
\begin{center} 
\includegraphics[height=7cm]{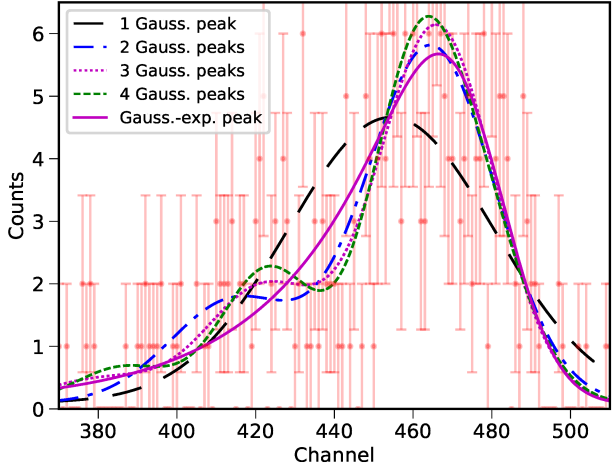} 
\caption{Profile curves corresponding to the likelihood maxima of the different models (1--4 Gaussian peaks and Gaussian-exponential peak).}
\label{fig:plot_compare}
\end{center}
\end{figure}

In this second example we consider the experimental data already presented in Sec.~\ref{sec:nested_fit} corresponding to the helium-like uranium $1s2p\ ^3\!P_2 \to 1s2s\ ^3\!S_1$ intrashell transition obtained from  a Bragg diffraction spectrometer equipped by a curved crystal \cite{Trassinelli2009b}.
As can be seen in Fig.~\ref{fig:plot_compare}, the experimental peak is statistically poor, quite broad and asymmetric. 
We do not know where this asymmetry comes from. 
Eventually, it might be related to the presence of several spectral components or from spectrometer's aberrations.
From the Bayesian analysis we would like i) to determine the most probable model that describes the data and ii) to determine the probability distribution of the main spectral component position, independently on the choice of the model.

For each model, we calculate with \texttt{Nested\_fit} the evidence, the probability distributions and the complexity using the same parameters as in the previous example except for the number of live points and the number of trial sets.
Here we use $K=2000$ live points and $N_{LPS}$ between 8 and 32 depending on the model.
For all models, we chose flat prior probability distributions.

\hfill

First we consider the simple case where we can have only Gaussian peaks, between one and four, with the same width $\sigma$, which we know to be a priori between 10 and 30 channels, and a flat background.
From these working hypotheses, we would like to determine which model is the most probable. i.e. how many peaks are present, and what is the position of the main peak.
To note, the model with four Gaussian peaks requires for any single trial set much more computation time than the single peak due to the presence of several high-value likelihood regions (see Fig.~\ref{fig:histo_peaks}).
This is in fact the practical reason why we consider a maximum of four components.
Similar examples have been presented in the past by Sivia \cite{Sivia1992,Sivia}. 
With respect to these works, here we consider the analysis of a statistically poor data set from a real experiment instead of a simulation, where we do not know the real nature of the spectra.

\begin{figure}
\begin{center} 
\includegraphics[height=7cm]{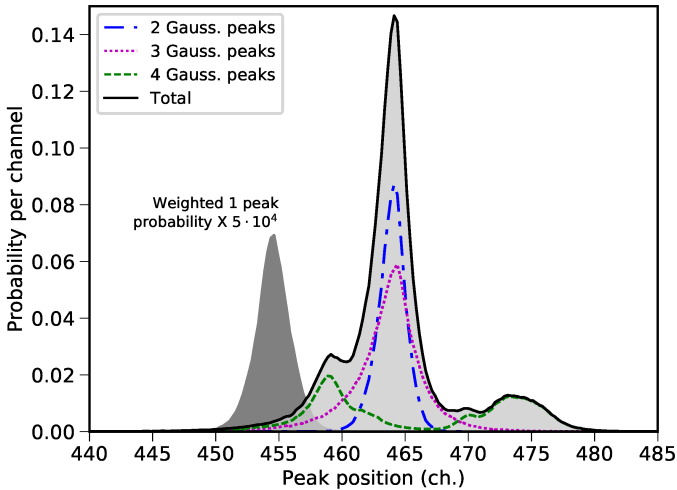} 
\caption{Probability distribution of the main peak position from the single probabilities of the models with one-to-four Gaussian peaks. 
For the single peak model, we magnify its weighted probability (in grey) to compare the distributions.}
\label{fig:plot_peak-pos}
\end{center}
\end{figure}

To visually compare the outcome of the different models, we present in Fig.~\ref{fig:plot_compare} the corresponding curves relative to each likelihood function maximum.
As it can be observe, the profile maxima are close to each other except for the single Gaussian peak profile.
In the particular case of the 4-peak model, two Gaussian component are unresolved (as suggested by Fig.~\ref{fig:plot_peak-pos}).

\begin{figure}
\begin{center} 
\includegraphics[width=0.9\columnwidth]{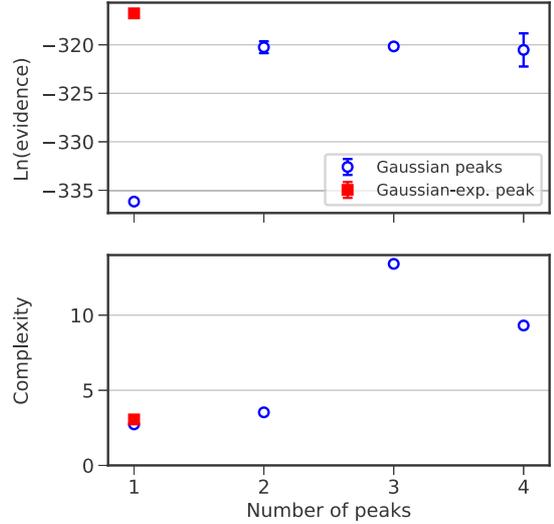} 
\caption{Evidence and complexity of the different considered models.}
\label{fig:evidence_comp}
\end{center}
\end{figure}

The quantitative results obtained from \texttt{Nested\_fit} are summarized in Table~\ref{tab:nasty} and Fig.~\ref{fig:evidence_comp} where we report values of the evidence (in the logarithmic scale), of the model complexity and the probability of the model (in the table only).
The model with a single Gaussian peak results to have a very low probability. 
From the results of the other hypotheses, we cannot clearly determine how many peaks are present because models with 2, 3 and 4 components have the same evidence (within the associated uncertainty).
As suggested by Trotta \cite{Trotta2008}, a criterium to choose between different models with similar evidence is the Bayesian complexity $\mathcal{C}$ value (see Sec.~\ref{sec:model_testing} and \ref{app:info}).
When different hypotheses have similar evidence values, we should choose the model with the lower value of $\mathcal{C}$ to favor once more simple models versus complex models, in agreement to the Ockham's razor principle.
In our case, the two-peak model is the favorite with a low complexity value, only slightly higher than the one-peak model complexity, and high probability.

\begin{table*}
\centering
\caption{For each model, the different probability values $P(\mathcal{M}| I)$ and related quantities are reported. In addition to the evidence value (in natural log), we report the model probability considering only Gaussian peaks ($P_\text{G. models}(\mathcal{M} |,I)$), a two-model probability with the two-Gaussian peak model as reference ($P_\text{Two-models}(\mathcal{M} |,I))$, the Bayesian complexity, the minimum value of reduced chi-square $\chi^2_{red}$ and the related probabilities ($P_\chi$ and  $P_F$) from $\chi^2$- and F-test.}
\label{tab:nasty}       
\begin{tabular}{l l p{2.0cm}  p{2.3cm} c c l l }
\hline
\hline
Model & $\ln E$ & $P_\text{G. models}$ & $P_\text{Two-models}$ & Complexity & $\chi^2_{red}$ & $P_\chi$ & $P_F$  \\			
\hline																	
\hline																	
1 Gauss. peak	 	& $	-336.17	\pm	0.19	$ & $	4.3	 \times 10^{-8}  $ & 	0.00001	\% &	2.7	&	0.8213	&	0.38885	\% &	10.1	\% \\
2 Gauss. peaks	 	& $	-320.25	\pm	0.61	$ & $	35.1	\%  $ &	--	&	3.5	&	0.7224	&	0.00081	\% &	--	 \\
3 Gauss. peaks	 	& $	-320.16	\pm	0.25	$ & $	38.2	\%  $ &	52.1	\% &	13.4	&	0.7009	&	0.00014	\% &	61.7	\% \\
4 Gauss. peaks	 	& $	-320.52	\pm	1.71	$ & $	26.6	\%  $ &	43.1	\% &	9.3	&	0.7022	&	0.00017	\% &	61.0	\% \\
\hline																	
1 Gauss.-exp. peak	 & $	-316.76	\pm	0.12	$ &  --		& 	97.0	\% &	3.1	&	0.7190	&	0.00060	\% &	48.1	\% \\
\hline
\hline
\end{tabular}
\end{table*}

If we are not interested to determine the number of peaks, but only to the main peak position component $\mu_0$, we can built the correspondent probability distribution $P(\mu_0| \{x_i, y_i\},I)$ from the output of the each model analysis.
As in Eq.~\eqref{eq:average}, we can build $P(\mu_0| \{x_i, y_i\},I)$ from the different $P(\mu_0| \{x_i, y_i\},\mathcal{M}_\ell,I)$ distributions using as weight the  model probabilities summarized in Table~\ref{tab:nasty}.
The final probability distribution of the main peak position (around channels 450--480) is presented in Fig.~\ref{fig:plot_peak-pos}. It is quite complex, with the presence of several maxima mainly due to the four-peak model contribution. 
These maxima correspond in fact to the high-likelihood regions visible in Fig.~\ref{fig:histo_peaks}.
Because of the low probability, the one-peak model does not contribute significantly to the final distribution.
As comparison its contribution is presented in Fig.~\ref{fig:plot_peak-pos} with a strong magnification factor.

\hfill

Alternatively to the presence of several Gaussian peaks, a valid hypothesis is the presence of some kind of aberration due to the spectrometer characteristics.  A spectrometer with cylindrically bent crystal in the Johann geometry is in fact used.
To take into account this possibility, we model the aberration effect by a line profile resulting from the convolution between a Gaussian and an exponential function \cite{Kalambet2011}.
As we can see in Fig.~\ref{fig:plot_compare}, the curve corresponding to the likelihood maximum reproduces well the data, with a maximum very close to the multi-Gaussian peaks models (with exclusion of the single peak model).
From Table~\ref{tab:nasty} and Fig.~\ref{fig:evidence_comp},  we can observe more quantitatively that the associated evidence is significantly higher than any other model. 
With respect to the two-Gaussian peaks, the probability for the Gaussian-exponential profile is in fact $P_\text{Two-models}(\mathcal{M} |,I) = 97.0\%$.
At the same time the associated complexity remains small, intermediate between single and double Gaussian peak models, indicating that, together with the high model probability value, the presence of an aberration as explanation of the asymmetry experimental data distribution is the most valid hypothesis.

To compare the results from the evidence analysis to classical goodness-of-fit tests such the $\chi^2$-test and the F-test, for each model, we find the minimal $\chi^2$ value using Minuit CERN library \cite{James1975}.
Due to the low statistics, we use a modified form of the $\chi^2$ derived from the Poisson distribution \cite{Baker1984}  in a homemade Fortran program (called \textit{Minuit\_fit}) that uses Minuit library.
For the $\chi^2$-test, we compute the probability $P_\chi$ for obtaining a higher value of the reduced chi-square $\chi^2_{red}$.
The low statistics causes low values of $\chi^2_{red}$ (significantly less than one) and accordingly small values of $P_\chi$ for all models.
For the F-test, we compute the probability $P_F$ for obtaining higher values of the $\chi^2_{red}$ ratio between the selected model and the two-Gaussian model considered as reference.
All values are reported in Table~\ref{tab:nasty}.
For completeness, in the table we report in addition the two-model probability $P_\text{Two-models}(\mathcal{M} |,I)$ computed from the evidence of the selected model and the two-Gaussian model. 
When only Gaussian profiles are considered, both $\chi^2$- and the F-test outcomes indicate that the single Gaussian peak model should be excluded, without a net preference of one of the multi-Gaussian peak models, similarly to the Bayesian analysis results. 
When the Gaussian-exponential model is considered in addition, the two approaches do not agree. 
In the comparison between Bayesian evidence values, the Gaussian-exponential has the highest probability.
In opposite, the F-test indicates an unclear preference between the two-Gaussian and the Gaussian-exponential models. 
To note that the F-test as well as the $\chi^2$ test are based on the probability for having a certain minimal value $\chi^2_{red}$ (i.e. maximum values of the likelihood function).
In opposite, $P_\text{Two-models}(\mathcal{M} |,I)$ consider the ensemble of the likelihood function over the parameter space, which includes much more information.
In addition we note that $P(\mathcal{M} |,I)$ are probabilities assigned to the different hypotheses calculated from the experimental data.
$P_\chi$ and $P_F$ are in contrary only probabilities linked to the statistical distribution of $\chi^2_{red}$ values that are are used as criteria to favor one model with respect another.
For this reason, they cannot be used to extract an average of a common parameter to the different models without selecting one precise model, which is possible from $P(\mathcal{M} |,I)$ values.

\hfill

In the previous paragraphs we show how evidence and complexity evaluations can help to determine the most plausible model to describe a set of data. 
In this specific example, we remember that we consider a strong assumption on the number of the possible Gaussian peaks to mainly limit the computational time. 
Other hypotheses could be considered but always taking into account our prior knowledge coming from previous experiments or general physical considerations.
Formally this prior knowledge should be included in the model prior probability that, once multiplied to the evidence, gives the final probability for the different models.
For this point, critics could be addressed about the objectivity. 
But again, the meaning of such dependency on the priors should be pragmatically be interpreted as a message saying that the data quality is not sufficient to correctly analyze the problem and choose among different hypotheses.
Nevertheless, this approach provide a well defined procedure to exclude unrealistic models with the comparison with the data via the evidence computation (as for the single-peak model)  or, via prior probabilities, models that are not consistent with our present knowledge of physics and simply common sense, on which our logic is based.

\section{Conclusions}\label{sec:conclusion}
The main intent of this article is to provide an useful starting point for the atomic physics community to use Bayesian methods for data analysis.
For this propose, we provide a very synthetic and basic introduction to Bayesian statistics.
We show how, from basic logic reasoning with requirement of consistency, a very general definition of probability can be constructed. 
This definition automatically implies the Bayes' theorem, which plays the central role for the prior probability inclusion.
From this approach, we see how posterior probabilities can be simply calculated as well as probabilities for different hypotheses.

To visualize the practical consequences of the use of these new concepts, we show two atomic spectra analysis examples.
In the first one we see how we can determine the presence or not of an unresolved spectral line.
In the second, more complex, we calculate the probability of different possible models (different number of peaks and shapes) and we see how to extract valuable information (the main peak position in our case) from equiprobable hypotheses.

For hypotheses testing, the calculation of the Bayesian evidence from the experimental data is essential.
Different methods are available in the literature to evaluate the Bayesian evidence. 
In this article we present in detail the nested sampling technique developed originally by J.~Skilling in 2004 based on a particular for of Monte Carlo sampling of the model parameter space.
We also present the newly developed program \texttt{Nested\_fit} based on such a method but with a new parameter exploration algorithm.
We show its capabilities and typical inputs and outputs.

As final general comment, we invite to use Bayesian methods to all cases where (i) hypotheses/models testing are involved and (ii) where constraints or prior knowledge on the model parameters are involved. 
As we saw, classical criteria based on goodness-of-fit tests can also be used.
In this case only the minimal values of $\chi^2$ are considered, and not their dependency on the entire possible range of parameter values. 
This can be dangerous for statistically poor data sets where small quantity of information is available.
In addition, from goodness-of-fit test outputs, the average of a parameter common to the different models is impossible to compute without selecting one precise model.
This issue is trivial with Bayesian statistics methods.


\section*{Acknowledgments}
First of all we would like to express our deep gratitude to Leopold M. Simons who introduced us to the Bayesian data analysis and without whom this work could not have been started.
We would like to sincerely thank also Nicolas Winckler for the numerous discussions about statistics and data analysis, and, together with Robert Grisenti, for the careful reading of the manuscript.
This work has been developed in the context of several experiments; we would like to thank all members of the Pionic Hydrogen, FOCAL and GSI Oscillation collaborations and the ASUR group at the Institute of NanoSicence of Paris for support and discussions.

\appendix

\section{Information and complexity} \label{app:info}
The gain of knowledge we obtain from the analysis of experimental data can be quantified in terms of information $\mathcal{H}$, in the Shannon sense  \cite{Shannon1948a,Shannon1948b}, comparing the posterior probability $P(\boldsymbol{a} | \{x_i, y_i\},I)$ with the prior probability $P(\boldsymbol{a}| I)$.
The information gain, in units of nat\footnote{\textit{nat} is the unit of information when the normal logarithm is used, similarly to the \textit{bit}, the unit where the base-2 logarithm is employed.}, is given by the so-called Kullback-Leibler divergence \cite{Kullback1951}
\begin{equation}
\mathcal{H} \equiv D_{KL} = \int P(\boldsymbol{a} | \{x_i, y_i\},I) \ln \left[ \frac {P(\boldsymbol{a} | \{x_i, y_i\},I)} {P(\boldsymbol{a}| I)} \right] d^D\boldsymbol{a}. \label{eq:information}
\end{equation}
Considering Eq.~\eqref{eq:bayes-exp}, $D_{KL}$ can be written as
\begin{equation}
D_{KL} = -\ln E + \int P(\boldsymbol{a} | \{x_i, y_i\},I) \ln L(\boldsymbol{a})  d^D\boldsymbol{a},
\end{equation}
which is nothing else that the negative logarithm of the evidence plus the average of the logarithmic value of the likelihood function.

From $D_{KL}$ an interesting quantity can be derived that provides an additional criterion to compare models: the \textit{Bayesian complexity} $\mathcal{C}$.
$\mathcal{C}$ is calculated from the difference between $D_{KL}$ and the ``expected surprise''  \cite{Trotta2008} from the data represented by the value $\hat D_{KL}$, with
\begin{equation}
\hat D_{KL} =  -\ln E +  \ln L(\boldsymbol{\hat a}),
\end{equation}
where $\boldsymbol{\hat a}$ usually correspond to the posterior parameter mean values, or other possible estimators (ex. the likelihood function maximum or the posterior distribution medians) depending on the details of
the problem\footnote{For multimode posterior probability distributions, the likelihood function maximum is more adapted. In fact the mean value can easily be far from the parameter region corresponding to high values of the likelihood function.}.
The complexity is then defined as \cite{Spiegelhalter2002,Trotta2008}
\begin{equation}
\mathcal{C} = -2 (D_{KL} - \hat D_{KL} ) 
= -2\left[ \langle \ln L(\boldsymbol{a}) \rangle -  \ln L(\boldsymbol{\hat a}) \right],
\end{equation}
where the symbol $\langle \ \rangle$ indicates the mean value.
$\mathcal{C}$ gives in practice a measurement of the number of parameters that the data can support for a certain model $\mathcal{M}$ for a defined parameter priors \cite{Spiegelhalter2002,Spiegelhalter2014}.

For equiprobable models (similar evidence values), the comparison of Bayesian complexity can be used to choose in favor to one hypothesis or another.
Considering  two different models $\mathcal{M}_1$ and $\mathcal{M}_2$ with $E_1 \approx E_2$ and different number of parameters $J_1 < J_2$, we can have to cases \cite{Trotta2008}:
\begin{description}
\item[$\boldsymbol{\mathcal{C}_1 < \mathcal{C}_2}$ :] The quality of the data is sufficient to measure the additional parameters of the more complicated model, but they do not improve its evidence by much. We should prefer model with less parameters.
\item[$\boldsymbol{\mathcal{C}_1 \approx \mathcal{C}_2}$:] The quality of the data is not sufficient to measure the additional parameters of the more complicated model and we cannot draw any conclusions as to whether extra parameters are needed.
\end{description}

\section{Theoretical uncertainty of the evidence calculation by nested sampling} \label{app:uncertainty}
The main uncertainty of the final evaluation of the evidence calculated by the nested sampling is, as stated by the author of this method J.~Skilling, related to the probabilistic nature of the terms $\Delta X_m$ in Eq.~\eqref{eq:evindence_app} \cite{Sivia,Skilling2006,Skilling2009,Veitch2010}. 
The choice of numerical integration of Eq.~\eqref{eq:evX} (rectangle method, trapezoidal rule, etc.) does not influence very much the final result.
Instead, the statistical glittering of $\Delta X_m$ in Eq.~\eqref{eq:evindence_app} introduces a significant error.

The interval values are calculated from $X_m = \prod_i^m t_i$ (Eq.~\eqref{eq:Xm}), where $t_i$ are the shrinking of the considered interval of $X$.
The statistical distribution of the shrinking values $t_i$ can be obtained from simple probabilistic considerations.
For each step $m$, the shrinking value is derived from the $\{ \xi_{m,k}\}$ values of $X$ that correspond to the $K$ considered live points.
The $K$ randomly sorted live points correspond the $K$ values $\{ \xi_{m,k}\}$ that are uniformly distributed in the interval $[0,X_m]$.
To pass to the $m+1$ step, we have to identify the maximum value of $\{ \xi_{m,k}\}$ to determine the shrinking factor $t_{m+1} = \max(\xi_{m,k}/X_m)$.
This correspond to find the maximum of $K$ values $\{x_k\}$ uniformly distributed in the interval $[0,1]$ (where $x_k = \xi_{m,k}/X_m$).

Considering a certain $x_{k'} = t$, the probability that all other values are less than $t$ is $\prod_{k \neq k'} P(x_k \in [0,t]) = t^{K-1}$.
Because this is valid for any $x_{k'} \in \{x_k\}$, we have 
\begin{equation} 
P(t = \max \{x_k\})=K t^{K-1}.
\end{equation}
This probability distribution has the following properties. 
The average and standard deviation of $\ln t$ are 
\begin{equation}
\langle \ln t \rangle = -\frac 1 K \quad \text{and} \quad \sigma_{\ln t} = \frac 1 K. \label{eq:lnt}
\end{equation}
From the above equation and Eq.~\eqref{eq:Xm}, we have 
\begin{equation}
\ln X_m = - \frac m K \pm \frac {\sqrt{m}} K. \label{eq:lnX}
\end{equation}
If the main value of $X_m$ is taken into account (as in Sec.~\ref{sec:nested}), we introduce an error of the order of $\sqrt{m}/K$ in the evidence evaluation via $\Delta X_m$.

As we see in Fig.~\ref{fig:step_weights}, not all $m$ steps contribute equally to for the final value of $E$. 
The calculated evidence is dominated by the region where the product $\mathcal{L}_m \Delta X_m$ is maximal.
The maximum position can correlate to the information gain $\mathcal{H}$ associated to the data (and the model) by Eq.~\eqref{eq:information}.

To estimate this position, we have to make some approximation.
Considering Eqs.~\eqref{eq:information}, \eqref{eq:dXvsda} and \eqref{eq:evX}, we have that the information in terms of $\mathcal{L}(X)$ is 
\begin{equation}
\mathcal{H} = \int_0^1 \frac {\mathcal{L}(X)} E \ln \left[ \frac {\mathcal{L}(X)} E \right] dX = \int_0^1 P(X) \ln P(X) dX.
\end{equation}
If we assume the extreme case of a likelihood function with a core with a constant value $\mathcal{L}(X) = \mathcal{ \hat L}$ for $X<\hat X$ 
and zero elsewhere \cite{Skilling2009}, we have that $E=\hat L \hat X$ and then $P(X) = 1/ \hat X$ for $X<\hat X$ and zero otherwise.
In this simple case we have
\begin{equation}
\mathcal{H} = \int_0^{\hat X} \frac 1 {\hat X} \ln \left( \frac 1 {\hat X} \right) dX = - \ln \hat X \label{eq:calc}
\end{equation}
and then $\hat X = e^{\mathcal{-H}}$ (see also Refs.~\citenum{Sivia,Skilling2006,Skilling2009,Veitch2010} for further considerations).

From Eqs.~\eqref{eq:lnX} and \eqref{eq:calc}, we see that the $m$ value associated to this region, the most influent region for the value of $E$, is $m=K \mathcal{H}$ and
\begin{equation}
\ln \hat X = \mathcal{H} \pm \sqrt{ \frac {\mathcal{H}} K}.
\end{equation}
The dominant uncertainty associated to the evidence is then
\begin{equation}
\delta (\ln E) \approx \delta \left[ \ln \left(\sum_m \Delta X_m\right) \right] \approx  \sqrt{ \frac {\mathcal{H}} K}. \label{eq:uncertaintyE}
\end{equation}

Many approximations in this evaluation have been done but the dependency of $\delta (\ln E) \propto 1/ \sqrt{K}$ emerges.
This dependency has been confirmed by computational studies \cite{Veitch2010} that also investigate the influence of the search algorithm parameters for the new live points in the nested sampling.

A more pragmatic and practical way to evaluate the accuracy of $E$, which is employed in \texttt{Nested\_Fit} program (see Sec.~\ref{sec:nested_fit}), is to calculate the evidence for different trials with different sets of live points and calculate then the average and the standard deviation of the different values of $\ln E$.
From the consideration above, this is in fact the natural estimation to study the uncertainty of $E$ \cite{Skilling2009,Chopin2010}.

\section*{References}

\bibliographystyle{apsrmp4-1}
\bibliography{nested_fit_rev}

\end{document}